\documentclass[genTeX]{nrc1}
\def\bmu{\mbox{\boldmath{$\mu$}}}
\def\bq{{\bf q}}
\def\br{{\bf r}}
\def\bS{{\bf S}}
\def\bJ{{\bf J}}
\newcommand{\STRUT}{\rule{0in}{2.5ex}}
\def\haf{\textstyle{1\over2}}

\begin{document}

\title{The Structure of Light Nuclei and Its Effect on Precise Atomic 
Measurements}
\author[J.\ L.\ Friar]{J.\ L.\ Friar}
\address{Theoretical Division, Los Alamos National Laboratory, Los Alamos, NM 
87545 USA.  \email{friar@lanl.gov}}
\shortauthor{Friar}
\maketitle

\begin{abstract}
My talk will consist of three parts: (a) what every atomic physicist needs to
know about the physics of light nuclei [and no more]; (b) what nuclear
physicists can do for atomic physics; (c) what atomic physicists can do for
nuclear physics. A brief qualitative overview of the nuclear force and
calculational techniques for light nuclei will be presented, with an emphasis on
debunking myths and on recent progress in the field. Nuclear quantities that
affect precise atomic measurements will be discussed, together with their
current theoretical and experimental status. The final topic will be a 
discussion of those atomic measurements that would be useful to nuclear physics.
\end{abstract}

\section*{Introduction}

\begin{quote} 
\em ....numerical precision is the very soul of science.....
\end{quote}

This quote\cite{darcy} from Sir D'Arcy Wentworth Thompson, considered by many to
be the first biomathematician, could well serve as the motto of this conference,
since precision is the {\it raison d'\^etre} of this meeting and this field. I
have always been in awe of the number of digits of accuracy achievable by atomic
physics in the analysis of simple atomic systems\cite{2S1S}.  Nuclear physics,
which is my primary field and interest, must usually struggle to achieve three
digits of numerical significance, a level that atomic physics would consider a
poor initial effort, much less a decent final result.

The reason for the differing levels of accuracy is well known: the theory of
atoms is QED, which allows one to calculate properties of few-electron systems
to many significant figures\cite{review}. On the other hand, no aspect of
nuclear physics is known to that precision.  For example, a significant part of
the ``fundamental'' nuclear force between two nucleons must be determined 
phenomenologically by utilizing experimental information from nucleon-nucleon
scattering\cite{2gen}, very little of which is known to better than 1\%. In
contrast to that level of precision, energy-level spacings in few-electron atoms
can be measured so precisely that nuclear properties influence significant
digits in those energies\cite{d-p}.  Thus these experiments can be interpreted
as either a measurement of those nuclear properties, or corrections must be
applied to eliminate the nuclear effects so that the resulting measurement tests
or measures non-nuclear properties. That is the purview of my talk.

The single most difficult aspect of a calculation for any theorist is assigning
uncertainties to the results.  This is not always necessary, but in calculating
nuclear corrections to atomic properties it is essential to make an effort. That
is just another way to answer the question,``What confidence do we have in our
results?'' Because it is important for you to be able to judge nuclear results 
to some degree, I have slanted my talk towards answers to two questions that
should be asked by every atomic physicist. The first is: ``What confidence
should I have in the values of nuclear quantities that are required to analyze
precise atomic experiments?'' The second question is: ``What confidence should I
have that the nuclear output of my experiment will be put to good use by nuclear
physicists?''

\section*{Myths of Nuclear Physics}

Every field has a collection of myths, most of them being at least partially
true at one time.  Myths propagate in time and distort the reality of the
present.  I have collected a number of these, some of which I believed in the
past.  The resolution of these ``beliefs'' also serves as a counterpoint to the
very substantial progress made in light-nuclear physics in the past 15 years, 
which continues unabated.

My myth collection includes:

$\bullet$  The strong interactions (and consequently the nuclear force) aren't
well understood, and nuclear calculations are therefore unreliable.

$\bullet$ Large strong-interaction coupling constants mean that perturbation
theory doesn't converge, implying that there are no controlled expansions in
nuclear physics.

$\bullet$  The nuclear force has no fundamental basis, implying that
calculations are not trustworthy.

$\bullet$  You cannot solve the Schr\"odinger equation accurately because of the
complexity of the nuclear force.

$\bullet$  Nuclear physics requires a relativistic treatment, rendering a
difficult problem nearly intractable.

All of these myths had some (even considerable) truth in the past, but today
they are significant distortions of our current level of knowledge.

\section*{The Nuclear Force}

Most of the recent progress in understanding the nuclear force is based on a
symmetry of QCD, which is believed to be the underlying theory of the strong
interactions (or an excellent approximation to it).  It is generally the case
that our understanding of any branch of physics is based on a framework of
symmetry principles. QCD has ``natural'' degrees of freedom (quarks and gluons)
in terms of which the theory has a simple representation.  The (strong) chiral
symmetry of QCD results when the quark masses vanish,  and is a more complicated
analogue of the chiral symmetry that results in QED when the electron mass
vanishes.  The latter symmetry explains, for example, why (massless or
high-energy) electron scattering from a spherical (i.e., spinless) nucleus
vanishes in the backward direction.

The problem with this attractive picture is that it does not involve the degrees
of freedom most relevant to experiments in nuclear physics: nucleons and pions.
Nevertheless, it is possible to ``map'' QCD (expressed in terms of quarks and
gluons) into an ``equivalent'' or surrogate theory expressed in terms of nucleon
and pion degrees of freedom.  This surrogate works effectively only at low
energy.  The small-quark-mass symmetry limit becomes a small-pion-mass symmetry
limit.  In general this (slightly) broken-symmetry theory has $m_{\pi} c^2 <<
\Lambda$, where the pion mass is $m_{\pi} c^2 \cong$ 140 MeV and $\Lambda \sim$
1 GeV is the mass scale of QCD bound states (heavy mesons, nucleon resonances,
etc.). The seminal work on this surrogate theory, now called chiral perturbation
theory (or $\chi$PT), was performed by Steve Weinberg\cite{QCD}, and many
applications to nuclear physics were pioneered by his student, Bira van
Kolck\cite{nQCD}. From my perspective they demonstrated two things that made an
immediate impact on my understanding of nuclear physics\cite{pc}: (1) There is
an alternative to perturbation theory in coupling constants, called ``power
counting,'' that converges geometrically like $(Q/\Lambda)^N$, where $Q \sim
m_{\pi}c^2$ is a relevant nuclear energy scale, and the exponent $N$ is 
constrained to have $N \geq 0$; (2) nuclear physics mechanisms are severely 
constrained by the chiral symmetry. These results provide nuclear physics with a
well-founded rationale for calculation.

This scheme divides the nuclear-force regime in a natural way into a long-range
part (which implies a low energy, $Q$, for the nucleons) and a short-range part
(corresponding to high energy, $Q$, between nucleons).  Since $\chi$PT only
works at low energies, we expect that only the long-range part of the nuclear
force can be treated by utilizing the pion degrees of freedom.  We need to
resort to phenomenology (i.e., nucleon-nucleon scattering data) to treat
systematically the short-range part of the interaction.

The long-range nuclear force is calculated in much the same way that atomic
physics calculates the interactions in an atom using QED.  The dominant
interaction is the exchange of a single pion, and is denoted $V_{\pi}$. Its
atomic analogue is one-photon exchange (containing the dominant Coulomb force).
Because it is such an important part of the nuclear potential, it is fair to
call $V_{\pi}$ the ``Coulomb force'' of nuclear physics.  Smaller contributions
arise from two-pion exchange (the analogue of two-photon exchange). There is
even an analogue of the atomic polarization force, where two electrons
simultaneously polarize their nucleus using their electric fields. The nuclear
analogue involves three nucleons simultaneously, and is called a three-nucleon
force\cite{3NF}. Although relatively weak compared to $V_{\pi}$ (a few percent),
it plays an important role in fine-tuning nuclear energy levels. The final
ingredient is an important short-range interaction (which must be determined by
phenomenology) that has no direct analogue in the physics of light atoms.

What are the consequences of exchanging a pion rather than a photon?  The
pseudoscalar nature of the pion mandates a spin-dependent coupling to a nucleon,
and this leads to a dominant tensor force between two nucleons.  Except for its
radial dependence, the form of $V_{\pi}$ mimics the interaction between two
magnetic dipoles, as seen in the Breit interaction, for example. Thus we have in
nuclear physics a situation that is the converse of the atomic case:  a dominant
tensor force and a smaller central force.  In order to grasp the difficulties
that nuclear physicists face, imagine that you are an atomic physicist in a
universe where magnetic (not electric) forces are dominant, and where QED can be
solved only for long-range forces and you must resort to phenomenology to
generate the short-range part of the force between electrons and nuclei.

Although this may sound hopeless, it is merely difficult. The key to handling
complexities is adequate computing power, and that became routinely available
only in the late 1980s or early 1990s. Since then there has been explosive
development in our understanding of light nuclei.  Underlying all of these
developments is an improved understanding of the nuclear force.  I like to
divide the history of the nuclear force into three distinct time periods.

{\bf First-generation} nuclear forces were developed prior to 1993. They all
contained the one-pion exchange force, but everything else was relatively crude.
The fits to the nucleon-nucleon scattering data (needed to parameterize the
short-range part of the nuclear force) were indifferent.

{\bf Second-generation} forces were developed in 1993 and in the years
following\cite{2gen}. They were more sophisticated and generally very well fit
to the scattering data.  As an example of how well the fitting worked, the
Nijmegen group (which pioneered this sophisticated procedure) allowed the pion
mass to vary in the Yukawa function defining $V_{\pi}$, and then fit that mass.
They also allowed different masses for the neutral and charged pions that were
being exchanged and found\cite{pi-mass}

\begin{equation}
m_{\pi^\pm} = 139.4(10)\, {\rm MeV} \, ,
\end{equation}
\begin{equation}
m_{\pi^{0}}\, = 135.6(13)\, {\rm MeV} \, ,
\end{equation}
both results agreeing with free pion masses ($m_{\pi^{\pm}} = 139.57018(35)$ MeV
and $m_{\pi^0} = 134.9766(6)$ MeV \cite{PDG}). It is both heartening and a bit
amazing that the masses of the pions can be determined to better than 1\% using
data taken in reactions that have no free pions!  This result is the best
quantitative proof of the importance of pion degrees of freedom in nuclear
physics.

{\bf Third-generation} nuclear forces are currently under development. These
forces are quite sophisticated and incorporate two-pion exchange, as well as
$V_{\pi}$.  All of the pion-exchange forces (including three-nucleon forces) are
being generated in accordance with the rules of chiral perturbation theory. One 
also expects excellent fits to the scattering data.  This is very much work in 
progress, but preliminary calculations and versions have already 
appeared\cite{3gen}.

\section*{Calculations of Light Nuclei}

Having a nuclear force is not very useful unless one can calculate nuclear
properties with it.  Such calculations are quite difficult. Until the middle
1980s only the two-nucleon problem had been solved with numerical errors smaller
than 1\%.  At that time the three-nucleon systems $^3$H and $^3$He were
accurately calculated using a variety of first-generation nuclear-force
models\cite{3N}.  Soon thereafter the $\alpha$-particle ($^4$He) was calculated
by my colleague, Joe Carlson, who pioneered a technique that has revolutionized
our understanding of light nuclei: Green's Function Monte Carlo
(GFMC)\cite{GFMC}.

The difficulty in solving the Schr\"odinger equation for nuclei is easily
understood, although it was not initially obvious to me.  Nuclei are best
described in terms of nucleon degrees of freedom. Nucleons come in two types,
protons and neutrons, which can be considered as the up and down components of
an ``isospin'' degree of freedom. If one also includes its spin, a single
nucleon has four internal degrees of freedom. Two nucleons consequently have 16
internal degrees of freedom, which is roughly the number of components in the
nucleon-nucleon force (coupling spin, isospin and orbital motion in a very
complicated way).  To handle this complexity one again requires fast computers,
and that is a fairly recent development.

The GFMC technique has been used by Joe Carlson and his collaborators to solve
for all of the bound (and some unbound) states of nuclei with up to 10 nucleons.
One of Joe's collaborators (Steve Pieper\cite{steve}) calculated that the
ten-nucleon Schr\"odinger equation requires the solution of more than 200,000
coupled second-order partial-differential equations in 27 continuous variables,
and this can be accomplished with numerical errors smaller than 1\%!  Their
results are very impressive.

Although the nucleon-nucleon scattering data alone can predict the binding
energy of the deuteron ($^2$H) to within about 1/2\%, the experimental binding
energy is used as input data in fitting the nucleon-nucleon potential. The
nuclei $^3$H and $^3$He are slightly underbound without a three-nucleon force,
and that force can be adjusted to remedy the underbinding. This highlights both
the dominant nature of the nucleon-nucleon force and the relative smallness of
three-nucleon forces, which is nevertheless appropriate in size to account for
the small discrepancies that result from using only nucleon-nucleon forces in
calculations.

The binding energy of $^4$He is then accurately predicted to within about 1\%.
The six-nucleon systems are also rather well predicted.  There are small
problems with neutron-rich nuclei, but only 3 adjustable parameters in the
three-nucleon force allow several dozen energy levels to be quite well
predicted.  I recommend that everyone peruse the impressive results of the GFMC
collaboration\cite{10A}.

We note finally that power counting can be used to show that light nuclei are
basically non-relativistic, and relativistic corrections are on the order of a
few percent.  Power counting is a powerful qualitative technique for determining
the relative importance of various mechanisms in nuclear physics.

\section*{What Nuclear Physics Can Do for Atomic Physics}

With our recently implemented computational skills we in nuclear physics can
calculate many properties of light nuclei with fairly good accuracy.  This is
especially true for the deuteron, which is almost unbound and is computationally
simple.  Although nuclear experiments don't have the intrinsic accuracy of
atomic experiments, many nuclear quantities that are relevant to precise atomic
experiments can also be measured using nuclear techniques, and usually with
fairly good accuracy.

What quantities are we talking about?  The nuclear length scale is $R \sim 1$ fm
$= 10^{-5}$ \AA.  The much larger atomic length scale of $a_0 \sim $ 1 \AA\
means that an expansion in powers of $R/a_0$ makes great sense, and a typical
wavelength for an atomic electron is so large compared to the nuclear size that
only moments of the nuclear observables come into play. An example is the
nuclear charge form factor (the Fourier transform of the nuclear charge density,
$\rho$), which is given by

\begin{equation}
F (\bq) = \int d^3 r \, \rho (\br) \, \exp ( i \bq \cdot \br ) 
\cong Z (1- \frac{\bq^2}{6}\langle r^2 \rangle_{\rm ch} + \cdots \, ) 
- \haf \, \bq^{\alpha} \bq^{\beta}\, Q^{\alpha \beta} + \cdots \, ,
\end{equation}
where $\bq$ is the momentum transferred from an electron to the nucleus,
$Q^{\alpha \beta}$ is the nuclear quadrupole-moment tensor, $Z$ is the total
nuclear charge, and $\langle r^2 \rangle_{\rm ch}$ is the mean-square radius of
the nuclear charge distribution. Since the effective $|\bq|$ in an atom will
be set by the atomic scales and consequently will be very small, these moments
should dominate the nuclear corrections to atomic energy levels.  If one then
uses $F$ to construct the electron-nucleus Coulomb interaction, one obtains

\begin{equation}
V_{\rm C} ( \br ) \cong - \frac{Z \alpha}{r} + \frac{2 \pi Z \alpha}{3} 
\langle r^2 \rangle_{\rm ch} \, \delta^3 (\br) - \frac{Q \alpha}{2 r^3} 
( 3 \, (\bS \cdot \hat{\br})^2 -\bS^2) + \cdots \, ,
\end{equation}
where $\bS$ is the nuclear spin operator and $Q$ is the nuclear quadrupole
moment (which vanishes unless the nucleus has spin $\geq 1$). The Fourier 
transform of the nuclear current density has a similar expansion

\begin{equation}
\bJ ( \bq ) = \int d^3 r \  \bJ (\br) \, \exp ( i \bq \cdot \br ) 
\cong -i \bq \times \bmu \, (1- \frac{\bq^2}{6} \langle r^2 
\rangle_{\rm M}\, + \, \cdots \, ) \; + \, \cdots \, ,
\end{equation}
where $\bmu$ is the nuclear magnetic-moment operator and $\langle r^2
\rangle_{\rm M}$ is the mean-square radius of the magnetization distribution.
The first term generates the usual atomic hyperfine interaction.

Electron-nucleus scattering is the primary technique used to determine those
nuclear moments of charge and current densities that are relevant to atomic
physics\cite{ingo}.  An exception is the measurement of the deuteron's
quadrupole moment [$Q = 0.282(19)$ fm$^2$] obtained by scattering polarized
deuterons from a high-Z target at low energy\cite{Qnuc}.  This result is
consistent with the molecular measurement [$Q = 0.2860(15)$
fm$^2$]\cite{Qhd1,Qhd2}, but its error is an order of magnitude larger. Although
there is no reason to believe that the tensor polarizability of the
deuteron\cite{tau} plays a significant role in the H-D (molecular) 
quadrupole-hyperfine splitting that was used to determine $Q$, that correction
was not included in the analysis. It was included in the analysis of the 
nuclear measurement.

I highly recommend the recent review by Ingo Sick\cite{ingo}, which contains
values of the charge and magnetic radii of light nuclei.  That review not only
contains the best values of quantities of interest, but discusses reliability
and many technical details for those who are interested. One qualitative result
from that review is important for the discussion below.  The errors of the
tritium ($^3$H) radii are about an order of magnitude larger than those of
deuterium.  Of all the light nuclei tritium is the most poorly known
experimentally, although the charge radius is relatively easy to calculate.

In addition to moments of the nuclear charge and current densities, various 
components and moments of the nuclear Compton amplitude can play a significant 
role. Examples are the (scalar) electric polarizability, $\alpha_E$, and the 
nuclear spin-dependent polarizability ($\sim \bS$). The latter term interacts 
with the electron spin to produce a contribution to the electron-nucleus 
hyperfine interaction.  There exists a recent calculation of the latter for 
deuterium\cite{d-nu}, and either calculations or measurements of $\alpha_E$ for
$^2$H\cite{d-pol,d-pol-g,d-pol-x}, $^3$H and $^3$He\cite{3pol}, and
$^4$He\cite{He4-x,He4-t}.

\section*{The Proton Size}

One recurring problem in the hydrogen Lamb shift is the appropriate value of the
mean-square radius of the proton, $\langle r^2 \rangle_{\rm p}$, to use in
calculations. Some older analyses\cite{HMW} disagree strongly with more recent
ones\cite{Simon}. As shown in Eqn.~(3), the slope of the charge form factor
(with respect to $\bq^2$) at $\bq^2$ = 0 determines that quantity. The form
factor is measured by scattering electrons from the proton at various energies
and scattering angles.

There are (at least) four problems associated with analyzing the charge 
form-factor data to obtain the proton size.  The first is that the counting
rates in such an experiment are proportional to the flux of electrons times the
number of protons in the target seen by each electron.  That product must be
measured.  In other words the measured form factor at low $\bq^2$ is ($a - b
\frac{\bq^2}{6} + \cdots$), where $b/a = \langle r^2 \rangle_{\rm p}$. The
measured normalization $a$ (not equal to 1) clearly influences the value and
error of $\langle r^2 \rangle_{\rm p}$. Most analyses unfortunately don't take
the normalization fully into account, and Ref.\cite{norm} estimates that a 
proper treatment of the normalization of available data could increase $\langle
r^2 \rangle_{\rm p}^{1/2}$ by about 0.015 fm and increase the error, as well. In
an atom, of course, the normalization is precisely computable.

Another source of error is neglecting higher-order corrections in $\alpha$
(i.e., Coulomb corrections). Ref.\cite{Coulomb} demonstrates that this increases
$\langle r^2 \rangle_{\rm p}^{1/2}$ by about 0.010 fm. A similar problem in
analyzing deuterium data was resolved in Ref.\cite{ST}. Another difficulty that
existed in the past was a lack of high-quality low-$\bq^2$ data.  The final
problem is that one must use a sufficiently flexible fitting function to
represent $F ( \bq)$, or the errors in the radius will be unrealistically low.
All of the older analyses had one or more of these flaws.

Most of the recent analyses\cite{Simon,Coulomb,fit} are compatible if the
appropriate corrections are made. An analysis by Rosenfelder\cite{Coulomb}
contains all of the appropriate ingredients, and he obtains $\langle r^2
\rangle_{\rm p}^{1/2}$ = 0.880(15) fm. There is a PSI experiment now underway to
measure the Lamb shift in muonic hydrogen, which would produce the definitive
result for $\langle r^2 \rangle_{\rm p}$ \cite{PSI,savely}. I fully expect the
results of that experiment to be compatible with Rosenfelder's result.
Extraction of the proton radius\cite{2loop} from the electronic Lamb shift is
now somewhat uncertain because of controversy involving the two-loop diagrams.
These diagrams are significantly less important in muonic hydrogen, where the
relative roles of the vacuum polarization and radiative diagrams are reversed.

\section*{What Atomic Physics Can Do for Nuclear Physics}

The single most valuable gift by atomic physics to the nuclear physics
community would be the accurate determination of the proton mean-square radius:
$\langle r^2 \rangle_{\rm p}$.  This quantity is important to nuclear theorists
who wish to compare their nuclear wave function calculations with measured
mean-square radii.  In order for an external source of electric field (such as a
passing electron) to probe a nucleus, it is first necessary to ``grab'' the
proton's intrinsic charge distribution, which then maps out the mean-square
radius of the proton probability distribution in the wave function: $\langle r^2
\rangle_{\rm wfn}$. Thus the measured mean-square radius of a nucleus, $\langle
r^2 \rangle$, has the following components:

\begin{equation}
\langle r^2 \rangle = \langle r^2 \rangle_{\rm wfn} + \langle r^2 \rangle_{\rm 
p}+ \frac{N}{Z} \langle r^2 \rangle_{\rm n} + \frac{1}{Z} \langle r^2 
\rangle_{\ldots} \, ,
\end{equation}
where I have included the intrinsic contribution of the N neutrons as well as
the Z protons, and $\langle r^2 \rangle_{\ldots}$ is the contribution of
everything else, including the very interesting (to nuclear physicists)
contributions from strong-interaction mechanisms and relativity in the nuclear
charge density\cite{czech}. Because the neutron looks very much like a 
positively charged core surrounded by a negatively charged cloud, its
mean-square radius has the opposite sign to that of the proton, whose core is
surrounded by a positively charged cloud. It should be clear from Eqn.~(6) that
$\langle r^2 \rangle_{\rm p}$ (which is much larger than $\langle r^2
\rangle_{\rm n}$) is an important part of the overall mean-square radius. Its
present uncertainty degrades our ability to test the wave functions of light
nuclei.

The next most important measurements are isotope shifts in light atoms or ions.
Since isotope shifts measure differences in frequencies for fixed nuclear charge
Z, the effect of the protons' intrinsic size cancels in the difference. This is
particularly important given the current lack of a precise value for the 
proton's radius. The neutrons' effect is relatively small and can be rather
easily eliminated, and thus one is directly comparing differences in wave
functions, or of small contributions from $\langle r^2 \rangle_{\ldots}$.
Isotope shifts are therefore especially ``theorist-friendly'' measurements,
since they are closest to measuring what nuclear theorists actually calculate.

Precise isotope-shift measurements have been performed for $^4$He -
$^3$He\cite{shiner} and for $^2$H - $^1$H (D-H)\cite{d-p}. A measurement of
$^6$He-$^4$He is being undertaken\cite{ANL} at ANL.  Gordon Drake has written
about and strongly advocated such measurements in the Li isotopes\cite{Li-IS}.
These are all highly desirable measurements. Because there are currently large
errors in the $^3$H (tritium) charge radius, in my opinion the single most
valuable measurement to be undertaken for nuclear physics purposes would be the
tritium-hydrogen ($^3$H - $^1$H) isotope shift.  An extensive series of
calculations using first-generation nuclear forces found $\langle r^2
\rangle_{\rm wfn}^{1/2}$ for tritium to be 1.582(8) fm, where the ``error''
is a subjective estimate\cite{radius}. This number could likely be improved by
using second-generation nuclear forces, although it will never be as accurate as
the corresponding deuteron value, which we discuss next.

The D-H isotope shift in the 2S-1S transition reported by the Garching 
group\cite{d-p} was

\begin{equation}
\Delta \nu = 670 \ \, 994 \ \, 334.64 (15) \ {\rm kHz} \, .
\end{equation}
\vspace*{-0.13in}
\hspace*{0.35in} $\uparrow$ \hspace{0.26in} $\uparrow$ \hspace{0.06in} 
$\uparrow$ \hspace{0.04in} $\uparrow$ \\ 
Most of this effect is due to the different masses of the two isotopes (and
begins in the first significant figure, indicated by an arrow). Nevertheless,
the precision is sufficiently high that the mean-square radius effect in the
sixth significant figure (second arrow) is much larger than the error.  The 
electric polarizability of the deuteron influences the eighth significant 
figure, while the deuteron's  magnetic susceptibility contributes to the tenth
significant figure.  It becomes difficult to trust the interpretation of the 
nuclear physics at about the 1 kHz level, so improving this measurement 
probably wouldn't lead to an improved understanding of the nuclear physics.

Analyzing this isotope shift and interpreting the residue (after applying all
QED corrections) in terms of the deuteron's radius leads to the
results\cite{iso} in Table 1. The very small binding energy of the deuteron
produces a long wave function tail outside the nuclear potential (interpretable
as  a proton cloud around the nuclear center of mass), which in turn leads to an
easy and very accurate calculation of the mean-square radius of the (square of
the) wave function. Subtracting this theoretical radius from the experimental
deuteron radius (corrected for the neutron's size) determines the effect of 
$\langle r^2 \rangle_{\ldots}$ on the radius. Although this difference is quite
small, it is nevertheless significant and half the size of the error in the
corresponding electron-scattering measurement. This high-precision analysis in 
Table 1 of the content of the deuteron's charge radius would have been
impossible without the precision of the atomic D-H isotope-shift measurement.
This measurement has given nuclear physics unique insight into small mechanisms
that are at present poorly understood\cite{MEC}.

\begin{table}[htb]
\centering
\topcaption{Theoretical and experimental deuteron radii for pointlike nucleons.
The deuteron wave function radius corresponding to second-generation nuclear
potentials and the experimental point-nucleon charge radius of the deuteron
(i.e., with the neutron charge radius removed) are shown in the first two
columns, followed by the difference of experimental and theoretical results. The
difference of the experimental radius with and without the neutron's size is
given last for comparison purposes\protect\cite{neutron}.}

\hspace{0.25in}

\begin{tabular}{|cc|c||c|}
\hline
$          \langle r^2 \rangle_{\rm wfn}^{1/2}\, ({\rm fm})$ & \STRUT
$ _{\rm exp}\langle r^2 \rangle_{\rm pt}^{1/2}\, ({\rm fm})$ &
$                             {\rm difference}\, ({\rm fm})$ & 
$     \Delta \langle r^2 \rangle_{\rm n}^{1/2}\, ({\rm fm})$ \\ \hline
1.9687(18) \rule{0in}{2.5ex}& 1.9753(10) & 0.0066(21) & -0.0291(7)\\ \hline

\end{tabular}
\end{table}

I have said very little in my talk about how information from hyperfine
splittings might provide insight into nuclear mechanisms. Partly this is due to
a lack of background on my part, and partly because the necessary calculations
haven't been performed. Karshenboim and Ivanov\cite{sav-nu} have compared
theoretical (QED only) and experimental results for various S-states in light
atoms, results that are expected to be accurate to roughly 1 part in $10^{8}$.
They find significant residual differences attributable to nuclear effects,
which range from tens of ppm to several hundred ppm. It is likely that these
differences contain interesting and useful nuclear information, in the form of
Zemach moments\cite{zemach} and spin-dependent polarizabilities. The latter are
related to the Drell-Hearn-Gerasimov sum rule\cite{DHG}, a topic of considerable
current interest in nuclear physics\cite{DHG-nuc}. Exploratory calculations are
underway.

\section*{Summary and Conclusions}

I hope that I have convinced you that nuclear forces and nuclear calculations in
light nuclei are under control in a way never before attained.  This progress
has been possible because of the great increase in computing power in recent
years. Many of the nuclear quantities that contribute to atomic measurements
have been calculated or measured to a reasonable level of accuracy, a level that
is improving with time.  Isotope shifts are valuable contributions to nuclear
physics knowledge, and are especially useful to theorists who are interested in
testing the quality of their wave functions for light nuclei. In special cases
such as the deuteron these measurements provide the only insight into the size
of small contributions to the electromagnetic interaction that are generated by
the underlying strong-interaction mechanisms. In my opinion the tritium-hydrogen
isotope shift would be the most useful measurement of that type. I am especially
hopeful for the success of the ongoing PSI experiment attempting to measure the
proton size via the Lamb shift in muonic hydrogen. The absence of a stable,
accurate proton radius has been particularly annoying.

\section*{Acknowledgements}
The work of J.\ L.\ Friar was performed under the auspices of the United States
Department of Energy. The author would like to thank Savely Karshenboim for an
invitation to this conference and for the opportunity to discuss the nuclear
side to precise atomic measurements.

\end{document}